\documentclass{emulateapj}

\slugcomment{}

\shorttitle{EXO 0748$-$676}

\shortauthors{Lin et al.}

\begin{document}
\title{The Incompatibility of Rapid Rotation with Narrow Photospheric
  X-ray Lines in EXO~0748$-$676}

\author{Jinrong Lin\altaffilmark{1}, Feryal \"Ozel\altaffilmark{2},
  Deepto Chakrabarty\altaffilmark{1}, and Dimitrios Psaltis\altaffilmark{2}}

\altaffiltext{1}{Department of Physics and Kavli Institute for
  Astrophysics and Space Research, Massachusetts Institute of
  Technology, Cambridge, MA 02139, USA; jinrongl, deepto@mit.edu}

\altaffiltext{2}{Department of Astronomy and Steward Observatory,
  University of Arizona, 933 N. Cherry Ave., Tucson, AZ 85721, USA;
fozel, dpsaltis@email.arizona.edu}

\begin{abstract}

X-ray observations of EXO~0748$-$676 during thermonuclear bursts
revealed a set of narrow ($\Delta \lambda /\lambda = 0.018$)
absorption lines that potentially originate from the stellar
photosphere. The identification of these lines with particular
atomic transitions led to the measurement of the surface
gravitational redshift of the neutron star and to constraints on
its mass and radius. However, the recent detection of 552~Hz
oscillations at 15\% rms amplitude revealed the spin frequency of
the neutron star and brought into question the consistency of such
a rapid spin with the narrow width of the absorption lines. Here,
we calculate the amplitudes of burst oscillations and the width of
absorption lines emerging from the surface of a rapidly rotating
neutron star for a wide range of model parameters.  We show that
no combination of neutron-star and geometric parameters can
simultaneously reproduce the narrowness of the absorption lines,
the high amplitude of the oscillations, and the observed flux at
the time the oscillations were detected. We, therefore, conclude
that the observed absorption lines are unlikely to originate from
the surface of this neutron star.
\end{abstract}

\keywords{stars:individual(EXO 0748-676) -- stars:neutron --
X-rays:binaries}

\section{Introduction}

The detection of gravitationally redshifted atomic spectral features
from the photospheres of neutron stars has long been recognized as the
optimal method for measuring neutron star compactness, hence
constraining the equation of state for ultradense matter \citep{bur63,
vpr82, lew93}. There were several reported detections of broad lines
in thermonuclear X-ray bursters during the 1980s \citep{waki84,
magnier89}. However, these observations were made using proportional
counters with modest spectral resolution and were difficult to
interpret in terms of the expected elemental abundance and ionization
profiles of burster atmospheres \citep{magnier89, mad90}. Their origin
is instead thought to be the instrumental Xe L edge.

More recently, \citet{cott02} reported the detection of three narrow
X-ray absorption lines in high-resolution X-ray grating spectra of the
burster EXO~0748$-$676 using {\em XMM-Newton}. They interpreted these
features as atomic transitions of highly ionized iron and oxygen at
the neutron star photosphere.  The significance of this detection was
bolstered by the fact that lines were all consistent with a single
gravitational redshift as well as with the observed ionization state
and temperature evolution of the burster atmosphere \citep{cott02,
chan05, rauch08}.  However, the narrowness of the lines
($\Delta\lambda/\lambda=0.018$) was surprising given the expectation
that most neutron stars in low-mass X-ray binaries are spinning at
millisecond periods, which should generally lead to broad line
profiles for most geometries \citep{ozel03, chan06}.  Subsequent
observations of the burster with the same instrument failed to confirm
these absorption lines \citep{cott08}, although this might be
explained by long-term variations in the spectral state of the source
\citep[e.g.,][]{krau07}.

At first, the puzzling narrowness of the detected lines seemed to be
explained by the discovery of a weak 45~Hz X-ray oscillation in the
summed power spectra of 38 thermonuclear bursts from the {\em Rossi
X-ray Timing Explorer (RXTE)}. This was interpreted as a spin
frequency sufficiently slow to give negligible spectral line
broadening \citep{vil04}. However, \citet{gall10} subsequently
detected more powerful 552~Hz oscillations in {\em RXTE} observations
of two 2007 bursts separated by 11 months.  The frequency and
amplitude of these oscillations were measured from single bursts,
without the need for summing power spectra.  The properties of these
burst oscillations allow a firm identification of 552~Hz as the spin
frequency \citep{gall10}, calling into question the photospheric
origin of the absorption lines.

Our work is motivated by the suggestion that a carefully chosen
geometry could still possibly preserve the interpretation of the
absorption lines observed by Cottam et al.\ (2002). The only way that
narrow photospheric absorption lines can be accommodated with a rapid
spin frequency is if the neutron star in EXO~0748$-$676 is viewed
nearly pole-on, so that the line-of-sight component of the surface
velocity is reduced. Such a configuration, however, will also
significantly suppress the amplitude of any burst oscillation
generated by a temperature inhomogeneity on the stellar surface. In
this paper, we explore this question by calculating the amplitudes of
burst oscillations and the width of absorption lines emerging from the
surface of a rapidly rotating neutron star.  Our aim is to determine
whether any combination of neutron-star and geometric parameters can
simultaneously reproduce the different X-ray observations of
EXO~0748$-$676: a 552~Hz oscillation at 15\% rms amplitude
\citep{gall10} and narrow absorption lines with $\Delta
\lambda /\lambda = 0.018$ at a gravitational redshift of $z=0.35$
\citep{cott02,vil04}.

\section{Models}

We begin by considering a neutron star rotating at 552~Hz. A
gravitational redshift of 0.35 corresponds to a compactness $2 G M /R
c^2 = 0.45$, where $M$ and $R$ are the neutron star mass and
radius. Because the width of lines from the neutron star surface
depends not only on the ratio $M/R$ but also on the radius itself, we
need to specify the $M$ and $R$ that yield the measured gravitational
redshift. We choose two pairs of masses and radii: $M=1.4~M_\odot$,
$R=9.2$~km and $M=2~M_\odot$, $R=13.1$~km.

We are interested in generating a pulse amplitude that is at least
as large as the observed value of 15\% (rms). This is because the
intrinsic amplitude can easily be reduced by a number of external
effects, but it cannot be amplified. In contrast, for the line
widths, we search for configurations that yield line widths that
are no broader than the observed value of 0.018. Additional
broadening mechanisms can exist, whereas there is no process that
can reduce the width beyond the intrinsic rotational broadening we
calculate here.

In our calculations, we model the exterior spacetime of the rotating
neutron star with a Schwarzschild metric but account for the
relativistic Doppler boosts and the time delays the photons experience
when they are emitted from the rapidly rotating surface. A comparison
of this approximation with the results obtained for numerical
spacetimes of rotating neutron stars show that the differences in the
lightcurves are not very large but that the Schwarzschild-plus-Doppler
approximation yields larger pulse amplitudes than do the exact
spacetimes \citep{cade07}.

The 552~Hz burst oscillations in EXO~0748$-$676 were detected in
the rise of the bursts, when the propagating burning front is
thought to create a short-lived temperature inhomogeneity on the
neutron star surface \citep[e.g.,][]{stro97}.  We model the hot
region as a single circular hot-spot with angular radius $\rho$
and assume that the rest of the neutron star is dark; this
assumption produces the largest rms amplitude in the burst
oscillation \citep{muno02}. The hot-spot is located at a
colatitude $\alpha$ from the ``north'' rotational pole of the
neutron star, taken to vary between 0$^\circ$ and 180$^\circ$. The
angle between the observer's line-of-sight and the rotation axis
is denoted by $\beta$, defined to lie between 0$^\circ$ and
90$^\circ$.

The light curves and the spectral line shapes also depend on the
spectra and the angular distribution of surface emission, which, in
turn, are shaped by the neutron star atmosphere \citep[see,
e.g.,][]{lon86, mad91}. Analyses of the observed burst spectra
indicate that they are usually Planckian \citep[e.g.,][]{gal08} and
can be adequately modeled as blackbodies. We, therefore, make this
assumption in our calculations. We describe the angular distribution
of photons emerging from the surface with a Hopf function
\citep{chan60,dedeo01}, which is appropriate for the
scattering-dominated neutron star atmospheres during a thermonuclear
burst.

With this setup, we calculated light curves from hot-spots radii of
$\rho=5^\circ - 45^\circ$ using the algorithm described by
\citet{muno02}, based on the technique outlined in \citet{pech83}
and \citet{ml98}. We then calculated the rms amplitudes of the
oscillations. The choice of small angular sizes of the hot-spot
maximizes the amplitude of the flux oscillations that can be
observed from the neutron star.

In the next section, we will also need to ensure that the
phase-averaged brightness of these hot-spots is sufficient to generate
the observed flux at the time when the burst oscillations were
detected. We, therefore, calculated the phase-averaged luminosity from
each spot using equation~(1) of \citet{psaltis00}.  In order to
convert this luminosity to a flux, we need an estimate of the distance
to EXO~0748$-$676. Even though spectroscopic estimates for the
distance to this source place it at $\gtrsim 7-9$~kpc
\citep{ozel06,boirin07}, we adopt here a very conservative lower limit
of 3~kpc in order to demonstrate that our results depend very weakly
on the assumed distance.

We calculated the spectral line broadening using the same formalism
and approach discussed in \citet{ozel03}. Relativistic Doppler boosts
cause an asymmetry in the spectral line profiles in addition to
broadening them, while the strong gravitational lensing effect alters
the relative contribution of surface elements with different
line-of-sight velocities to the line profile. We take the input
spectrum to be a Gaussian line with negligible intrinsic width. We
define $\Delta\lambda$ as the full-width half-maximum of the output
spectral feature and the broadening of the spectral line as
$\Delta\lambda / \lambda$. Note that the absorption lines were
detected in the cooling tail of the bursts, when the emission comes
from the entire surface of the neutron star rather than a localized
hot spot.  Thus, the spectral width will depend on the observer's
inclination $\beta$ but not on the hot-spot colatitude $\alpha$.

\begin{figure}[t]
\epsscale{1} \plotone{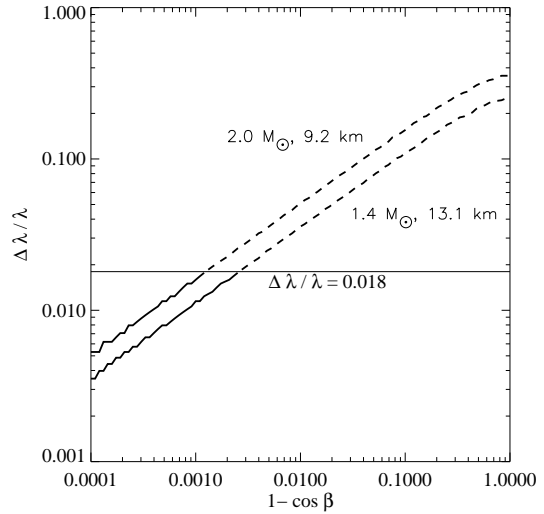} \caption {The minimum fractional
width $\Delta \lambda/\lambda$ of an intrinsically narrow atomic
line emitted from the entire surface of a 552~Hz rotating neutron
star as a function of $1-\cos \beta$, where $\beta$ is the angle
between the rotation axis of the neutron star and the
line-of-sight to an observer at infinity. The two curves show the
result for two neutron stars with different masses and radii that
correspond to a surface gravitational redshift of $z=0.35$. The
solid portion of the curves indicates where the $\beta$ value is
consistent with the observed line width limit ($\Delta
\lambda/\lambda \leq 0.018$; horizontal line) reported by
\citet{cott02}. Only a small range of $\beta$ values is consistent
with this limit.} \label{fig:width}
\end{figure}

\newpage

\section{Results}

We present here the constraints the two observations impose on the
geometry of the system for the value of the gravitational redshift of
$z=0.35$ inferred by \citet{cott02}. Figure~\ref{fig:width} shows the
fractional width of an intrinsically narrow atomic line emitted from
the entire surface of the neutron star, as a function of the cosine of
the observer's inclination $\beta$. As expected, the line broadening
due to rotation scales roughly as the non-relativistic expression
\begin{eqnarray}
\frac {\Delta \lambda} {\lambda} &\simeq& \frac {2
\Omega R}{c} \sin \beta \nonumber \\ &=&
0.21 \left(\frac{\nu_{\rm s}}{552~{\rm Hz}}\right)
\left(\frac{R}{9.2~{\rm km}}\right)\left(1-\cos^2\beta\right)^{1/2}.
\end{eqnarray}
As the angle $\beta$ decreases (and hence $1-\cos\beta$ also
decreases), the projection of the rotation velocity along the line of
sight also decreases, causing the rotational broadening of the line to
be substantially reduced. For a neutron star of 1.4~M$_\odot$, the
observed upper limit of $\Delta \lambda/\lambda\le 0.018$
\citep{vil04} on the fractional width of the lines reported by
\citet{cott02} can be satisfied only if $\cos\beta\gtrsim (1- 2.6
\times 10^{-3})$ or, equivalently, if the observer's inclination is
within $4.1^\circ$ from the rotation axis of the neutron star.

\begin{figure}[t]
\epsscale{1} \plotone{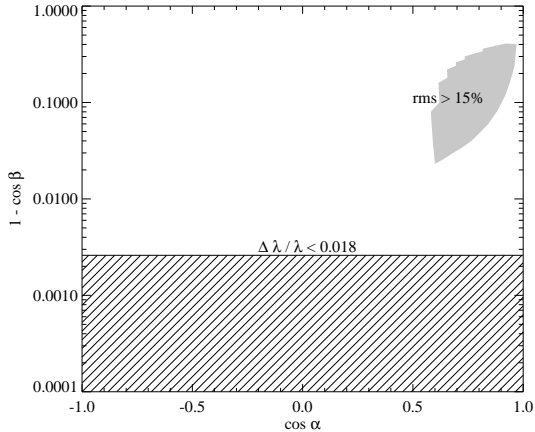}
\caption{Constraints on the inclination angle $\beta$ of an observer at
infinity and on the colatitude $\alpha$ of a localized emission region
on the neutron star surface responsible for the burst oscillations
during the rise phase of X-ray bursts for EXO 0748$-$676.  The
rectangular hatched region shows the limit on the observer's
inclination imposed by the narrow widths of the absorption lines
reported by \citet{cott02}. The gray shaded region shows the
constraints imposed by the 15\% (rms) burst oscillation amplitude
reported by \citet{gall10} and the phase-averaged flux observed during
the oscillations. Both calculations were performed for a hot-spot
radius of 5$^\circ$ on a neutron star with a mass of 1.4~$M_\odot$ and
a radius that corresponds to a surface gravitational redshift of
$z=0.35$. There are no pairs of inclination and colatitude angles that
are simultaneously consistent with both constraints.}
\label{fig:contours}
\end{figure}

The posterior chance probability that we are observing the neutron
star in EXO~0748$-$676 within 4.1$^\circ$ of its rotation axis is
exceedingly small. Assuming a random distribution in the orientation
of observers in the sky, i.e., a distribution that is flat in
$\cos\beta$, leads to a chance probability of $\le 2.6 \times
10^{-3}$. More importantly, as we show below, such a face-on
orientation of the system is inconsistent with the $\ge 15$\%
amplitudes of burst oscillations reported by \citet{gall10} in the
same source.

\begin{figure}[t]
\epsscale{1} \plotone{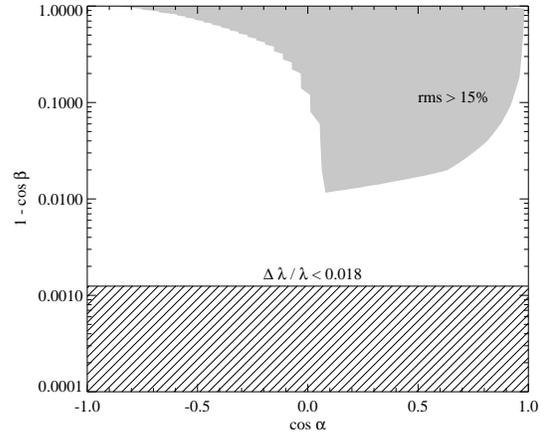} \caption {Same as Figure~\ref{fig:contours}
for a neutron star with a mass of 2~$M_\odot$ and a radius that
corresponds to a surface gravitational redshift of $z=0.35$.}
\label{fig:contours_2M}
\end{figure}

Figure~\ref{fig:contours} shows the constraint on the inclination
angle $\beta$ of the observer and the colatitude $\alpha$ of the hot
emitting region that gives rise to the burst oscillations during the
rise phase of X-ray bursts. This constraint is obtained by imposing
two requirements~\citep[see][]{psaltis00}: The rotating hot spot
should produce oscillations with an rms amplitude of at least 15\% and
the phase averaged flux in the 2.5$-$25~keV range should be larger
than $3 \times 10^{-10}$~erg~s$^{-1}$~cm$^{-2}$, observed at the time
when the oscillations were detected \citep{gall10}. The first
requirement constrains the observer's inclination $\beta$ and the
colatitude of the spot $\alpha$ to be away from the rotational pole of
the star. The second requirement excludes nearly antipodal
orientations between the observer and the hot-spot, in which case the
spot is almost completely hidden from view.

These two requirements can be met in the light gray shaded region,
i.e., only when the inclination of the observer is $\beta>11^\circ$,
such that $(1-\cos\beta)>2\times 10^{-2}$. If the inclination of the
observer is smaller than this value or the colatitude of the emitting
region is close to zero, then the amplitude of the burst oscillations
is reduced below the detected value. If, on the other hand, the
colatitude of the hot-spot is $\gtrsim 55^\circ$, so that $\cos \alpha
\lesssim 0.6$, then the rms amplitudes can be very large but the
phase-averaged flux goes to zero. In our calculations, we assume that
all of the emission originates from the hot spot. Taking into account
any emission from the rest of the neutron star, at a lower
temperature, may help increase the observed flux, but this would also
significantly reduce the pulse amplitude that can be detected.

The hatched region shows the area of the parameter space that is
consistent with the width of the narrow lines reported by
\citet{cott02}. As discussed earlier, this places a constraint only on
the observer's inclination $\beta$ because the entire neutron star is
expected to emit during the cooling tails of the bursts when the lines
were detected. Figure~\ref{fig:contours} demonstrates, therefore, that
the narrow width of the reported atomic lines is inconsistent with the
observed amplitudes of the burst oscillations for any emission
geometry and observer orientation with respect to the rotation axis of
the neutron star.

\begin{figure}[t]
\epsscale{1} \plotone{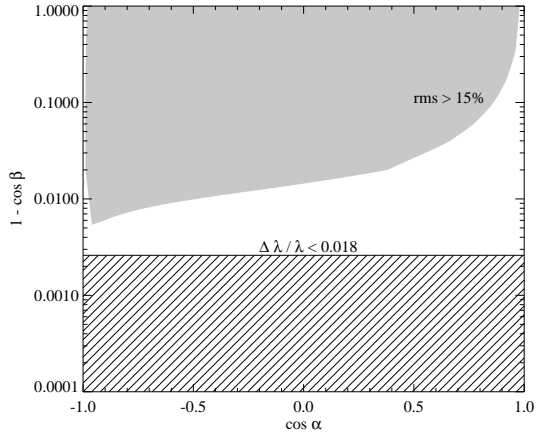} \caption {Same as
Figure~\ref{fig:contours} for a hot-spot radius of 45$^\circ$.}
  \label{fig:contours_45}
\end{figure}

The particular constraints shown in Figure~\ref{fig:contours}
depend on the mass of the neutron star as well as the angular size
of the hot-spot. Increasing the mass while keeping the redshift
the same results in a neutron star with a larger radius. This
causes the constraint imposed by line broadening to be even
stricter (see Fig.~\ref{fig:width} and eq.~[1]), and the observed
flux for a given angular spot size to be larger.
Figure~\ref{fig:contours_2M} shows these effects for a
2.0$~M_\odot$ neutron star and demonstrates that even though the
individual constraints change, there is still no part of the
parameter space that is consistent with both requirements.

Increasing the size of the hot-spot does not change this conclusion
either. We show in Figure~\ref{fig:contours_45} the allowed regions of
the parameter space calculated for a hot-spot radius of
45$^\circ$. Even though a larger range of hot-spot colatitudes can
generate the observed flux, the larger size of the hot-spot
substantially reduces the amplitude of oscillations (especially at the
antipodal orientations between the hot-spot and the observer), thus
limiting further the observer's inclination.  The combination of these
two effects leaves again no region of the parameter space that is
consistent with the width of the atomic lines.

We also considered possible uncertainties in the timing
measurement. Our calculations show that no region of the parameter
space is allowed unless the rms amplitude is smaller than 7$\%$. Such
a small value of the amplitude is excluded by the timing observations
at more than a 4$\sigma$ level. Along the same lines, we explored the
possibility that the detected 552~Hz burst oscillation is at the
second harmonic of the spin frequency of the neutron star. We verified
that there is still no combination of parameters that is consistent
with both requirements, even without including the further geometrical
constraints that are necessary to make the second harmonic stronger
than the fundamental.

Finally, our conclusion depends rather weakly on the identification of
the reported spectral features with particular atomic lines and hence
on the assumed gravitational redshift from the surface of the neutron
star. \citet{rauch08} identified the observed spectral features in
EXO~0748$-$676 with a different set of atomic transitions and
concluded that the gravitational redshift from the neutron star
surface is $z=0.24$. We repeated the above calculations for this value
of the redshift and found again that no region of the parameter space
is simultaneously consistent with both observations.
We, therefore, conclude that the lines reported by \citet{cott02}
are unlikely to originate from the photosphere of this neutron
star. As a result, any inference of the neutron star mass and
radius based on this redshift measurement \citep[e.g.,][]{ozel06}
is not tenable.

\acknowledgements J.L. and D.C. acknowledge support from the NASA
Astrophysics Data Program. F\"O acknowledges support from NSF
grant AST 07-08640 and Chandra Theory grant TMO-11003X. DP was
supported by the NSF CAREER award NSF 0746549 and Chandra Theory
grant TMO-11003X.


\end{document}